\documentclass[twocolumn,showpacs,preprintnumbers,amsmath,amssymb]{revtex4}
\usepackage{amsmath}
\usepackage{psfrag}
\usepackage{color}
\usepackage{latexsym,amsfonts}
\usepackage{graphicx}
\providecommand{\href}[2]{#2}   

\definecolor{Blue2}{rgb}{0.,0.,0.8125}
\definecolor{Brown3}{rgb}{0.625,0.25,0.}
\definecolor{Cyan4}{rgb}{0.,0.56,0.56}
\definecolor{Green4}{rgb}{0.,0.56,0.}
\definecolor{LtBlue}{rgb}{0.27,0.42,0.52}
\definecolor{Magenta4}{rgb}{0.5625,0.,0.5625}
\definecolor{Red2}{rgb}{0.8125,0.,0.}
\newcommand{\be}{\begin{equation}}
\newcommand{\ee}{\end{equation}}
\newcommand{\bee}{\begin{eqnarray}}
\newcommand{\eee}{\end{eqnarray}}

\begin{document}
\title{\bf Effect of the Coulomb interaction in  $A(d,p)$ fragmentation}%

\author{A.P.~Kobushkin}
 \email{kobushkin@bitp.kiev.ua}
\affiliation{
        Bogolyubov Institute for Theoretical Physics,03680, Kiev, Ukraine
}%
\author{Ya.D.~Krivenko-Emetov}
\affiliation{
        Institute for Nuclear Research, National Academy of Sciences, 03680  Kiev, Ukraine}%

\date{}      



\begin{abstract}
In the framework of Glauber-Sitenko model we calculate contribution of Coulomb interaction in cross-section of $A(d,p)$ reaction at high energy and zero angle. It is demonstrated that such effectsignificantly increases the differential cross section only at peak, where the proton momentum $p$ is near half of the deuteron momentum  $p_d$ in lab. frame, $p \sim \frac12 p_d$. The Coulomb interaction do not change the results in the high momentum region, where quark effects should be taken into account.
\end{abstract}

\maketitle

\section{\label{sec:Introduction}Introduction}
For several decades considerable efforts have been done to investigate structure of the deuteron in a wide inter-nuclear region, from that where description is given in terms of nucleons and mesons to that where quarks and gluons should be explicitly used for the deuteron description. Reactions with ``hadron probe'', due to its high luminosity, play important role in such studies. Here we mention only the  $A(d,p)$ break-up at zero proton angle \cite{BUPCS1,BUPCS2,BUPT20Perdrisat,BUPT20Ableev,BUPT20Punjabi,BUPT20Ableev90,BUPT20Nomifilov,BUPT20Azgirey,BUPKAPPACheunh,BUPKAPPADzikowski,BUPKAPPAKuehn} and the elastic $pd$-scattering at $180^\circ$ in c.m. frame \cite{EBSCS,EBST20,EBSKAPPA}. For both reactions detailed data on differential cross section and polarization observables, tensor analyzing power, $T_{20}$, and polarization transfer, $\kappa_0$, were obtained.

This data show significant deviation of the observables (cross-section,  $T_{20}$ and $\kappa_0$) from theoretical calculations, which cannot be removed by multiple scattering effects without modification of the short-range part of the deuteron wave function \cite{Kobushkin98}. Such modification comes from Pauli principlå on a level of constituent quarks \cite{GKS,GlozmanKuchina}. As a result, the short range part of the deuteron wave function includes, apart from the $pn$ component, new components, $\mathrm{NN}^\ast$, $\mathrm{N^\ast N}$ and $\mathrm{N^\ast N^\ast}$, which cannot be reduced to the $pn$ configuration. Because the main contribution of such states comes from the lowest resonances which have negative parity, the modification generates effective $P$-wave in the deuteron which drastically changes behaviour of the observables at high momentum region.

In the framework of such approach a good description of all break-up data was obtained \cite{Kobushkin98}. Nevertheless one important problem was not yet discussed. The data were measured at zero proton angle, $\theta_p=0^\circ$, and Coulomb interaction might, in principle, give a sizable contribution in the observables. In the present analysis, we account for Coulomb interaction by adopting the Ahkiezer and Sitenko approach \cite{AkhiezerSitenko} \footnote{For the further applications of this model see, e.g., \cite{EPS}.} to inclusive $^{12}\mathrm C(d,p)$ break-up and find that it significantly increases the cross section only at peak when the proton momentum $p$ is near half of the deuteron momentum $p_d$ in the lab. frame, $p\sim \frac12 p_d$.

The paper is organized as follows. In Sect.~\ref{sec:genral} we formulate the general formalism. We get expression for the Coulomb corrected amplitude (Sect.~\ref{sec:1.1}), give remarks to strong interacting part of the amplitude (Sect.~\ref{sec:1.2}) and obtain the general expression for the invariant differential cross section of the $(d,p)$ break-up at zero proton angle (Sect.~\ref{sec:1.3}). The procedure of numerical calculations is discussed in Sect.~\ref{sec:numerical}. In Sect.~\ref{sec:summ} we compare our calculations with experiment and  give a summary.
\section{Formalism \label{sec:genral}}
\subsection{Coulomb corrected amplitude \label{sec:1.1}}
We begin by considering general formalism for the Coulomb correction to cross-section of the deuteron break-up $A(d,p)$ at the proton longitudinal momentum $p_3\sim \frac12 p_d$ and small transverse momentum $p_\perp$. In this case it is enough to take into account the elastic contribution only.

Neglecting the Coulomb interaction the amplitude for the elastic break-up reads
\be
\begin{split}
&F^{\mathrm{\;str.}}(\mathbf p_\perp,p_3,\mathbf Q_\perp)=\frac{ip_d}{2\pi}\int d^2Be^{i\mathbf Q_\perp\mathbf B}\times \\
&\times
\int d^3r \psi^\ast_\mathbf{\,k}(\mathbf r)\psi_0(\mathbf r)\left[1-e^{i\chi_\mathbf{str}(\mathbf b_p,\mathbf b_n)}\right]= \\
&=\frac{ip_d}{2\pi}\int d^2Be^{i\mathbf Q_\perp\mathbf B}
\int d^3r \psi^\ast_\mathbf{\,k}(\mathbf r)\psi_0(\mathbf r)\times \\
&\times\left[
\Gamma_n(\mathbf b_n)+\Gamma_p(\mathbf b_p)-\Gamma_n(\mathbf b_n)\Gamma_p(\mathbf b_p)
\right]= \\
&=F^{\mathrm{\;str.}}_n(\mathbf p_\perp,p_3,\mathbf Q_\perp)+F^{\mathrm{\;str.}}_p(\mathbf p_\perp,p_3,\mathbf Q_\perp)-\\
&-F^{\mathrm{\;str.}}_{np}(\mathbf p_\perp,p_3,\mathbf Q_\perp)
,
\end{split}
\label{eq:1}
\ee
where $\psi_0(\mathbf r)$ and $\psi_\mathbf{\,k}(\mathbf r)$ are the wave functions of the deuteron and the  final proton-neutron system and $\Gamma_p(\mathbf b_p)$ and $\Gamma_n(\mathbf b_n)$ are the profile functions for the proton and the neutron; $\mathbf B=\frac12(\mathbf b_p + \mathbf b_n)$ and $\mathbf r_\perp=\mathbf b_p - \mathbf b_n$, with $\mathbf b_p$ and  $\mathbf b_n$ are impact parameters for the proton and the neutron; $\mathbf p_d=(0,0,p_d)$ is the deuteron momentum, $\mathbf Q_\perp$ is the momentum transferred to the final proton-neutron system. The definition of the relative momentum between the final proton and neutron, $\mathbf k$, needs a special comment. In the non-relativistic case it is defined as $\mathbf k=(\mathbf p_\perp-\frac12\mathbf Q_\perp,p_3)$. For the relativistic deuteron, $p_d\gg m_d$, we define it by boosting along $z$~axis to a frame where the total longitudinal momentum of the two-nucleon system is zero,
\bee
&p_3^\ast + n_3^\ast =0, \qquad \mathbf{p_\perp^\ast=p_\perp},\qquad \mathbf{n_\perp^\ast=n_\perp},\label{new_frame}\\
&\mathbf k_\perp=\frac12(\mathbf{p_\perp -n_\perp})=\mathbf p_\perp-\frac12\mathbf Q_\perp,\ k_3=\frac12(p_3^\ast -n_3^\ast),\label{rel_mom}
\eee
where $\mathbf p^\ast$ and $\mathbf n^\ast$ are momenta of the final proton and neutron in the new frame; $\mathbf p$ and $\mathbf n$ are the same momenta in the lab. frame. Of course, such definition can be accepted only near the region $p_3 \sim \frac12 p_d$. Assuming that the transverse motion is non-relativistic one simply gets
\be\label{rel_mom_expl}
k_3=\frac{\sqrt{M^2+\frac12p_d^2}}{E_d}\,\widetilde p, \qquad\text{where} \qquad p_3=\frac12 p_d +\widetilde p.
\ee
The profile function is given by
\be
\begin{split}
\Gamma(\mathbf b)=&\frac1{2\pi i p_N}\int d^2le^{-i\mathbf{l b}} f_N(q)=\\
=&\frac{(1-i\rho_N)\sigma_N}{4\pi \beta_N^2}e^{-\frac12 b^2/\beta_N^2},
\end{split}
\ee
where 
\be
f_N(l)=\frac{(i+\rho_N)p_N\sigma_N}{4\pi}e^{-\frac12\beta^2_Nl^2}
\label{NAampl}
\ee
with $\beta_N^2$ is the slope parameter for the nucleon-nucleus scattering, $\sigma_N$ is the total cross section and $\rho_N$ is the ratio of real to imaginary parts of the amplitude. 
%
Laler on we will use a natural approximation $\sigma_p=\sigma_n\equiv \sigma$, $\beta_p=\beta_n\equiv \beta$ and $\rho_N=0$.

Following the prescription of Ref.~\cite{GlauberMatthiae} the Coulomb interaction is included by adding the Coulomb, $\chi_\mathrm{c}(b_p)$, and screening Coulomb, $\chi_\mathrm{scr}$,  phase shifts to strong interacting phase function $\chi_\mathbf{str}(\mathbf b_p,\mathbf b_n)$, i.e.
\be
\begin{split}
\Gamma(\mathbf b_p,\mathbf b_n)&=1-e^{i\chi_\mathrm{str}(\mathbf b_p,\mathbf b_n)}\to \\
&\to 1-e^{i[\chi_\mathrm{str}(\mathbf b_p,\mathbf b_n)+\chi_\mathrm{scr}+\chi_\mathrm{c}(b_p)]}=\\
&=\Gamma(\mathbf b_p,\mathbf b_n)+ e^{i\chi_\mathrm{scr}}
\left[
e^{-i\chi_\mathrm{scr}}-e^{i\chi_\mathrm{c}(b_p)}
\right]\times\\
&\times\left[1-\Gamma(\mathbf b_p,\mathbf b_n)\right],
\end{split}
\label{eq:1.a}
\ee
where the Coulomb and screening Coulomb phase shifts are given by
\bee
\chi_\mathrm{c}(b_p)&=&
\frac{2Z\alpha}{v_p}\ln p_3 +\frac{4\pi Z \alpha}{v_p}\left[
\ln b_p\int_0^{b_p} T_\mathrm{c}(b')b'db' +\right.\nonumber\\
&&\left. +\int_{b_p}^\infty T_\mathrm{c}(b')\ln b'b'db'\right] \equiv\chi_0+\widetilde\chi_\mathrm{c}(b_p),
\label{eq:2}\\
\chi_\mathrm{scr}&=&-\frac{2 Z \alpha}{v_p}\ln 2 p_3R_\mathrm{scr},\label{eq:3}
\eee
respectively. Here $T_\mathrm{c}(b)=\int \rho_\mathrm{c}(r)dz$ is the thickness function corresponding to the nucleon charge distribution, $\rho_\mathrm{c}(r)$, normalized by $\int \rho_\mathrm{c}(r) d^3r=1$; $v_p$ is the proton velocity, $R_\mathrm{scr}$ is atomic screening radius, $Z$ is the atomic number and $\alpha\approx1/137$ is the fine structure constant. 

After that the Coulomb amplitude reads
\be
\begin{split}
&F^{\mathrm{\;c}}(\mathbf p_\perp,p_3,\mathbf Q_\perp)=e^{i\chi_\mathrm{scr}}\frac{ip_d}{2\pi}\int d^2Be^{i\mathbf Q_\perp\mathbf B}
\int d^3r \times\\
&\times \psi^\ast_\mathbf{\,k}(\mathbf r)\psi_0(\mathbf r)
\left\{\left[e^{-i\chi_\mathrm{scr}}-e^{i\chi_\mathrm{c}(b_p)}\right]\right\}\times\\
&\times\left(1-\Gamma_p-\Gamma_n+\Gamma_p\Gamma_n\right)=\\
&=F^{\mathrm{\;c}}_\mathrm{dis}-F^{\mathrm{\;c}}_p-F^{\mathrm{\;c}}_n+F^{\mathrm{\;c}}_{pn}.
\end{split}
\ee

Now let us consider contribution of different terms of this expression.
\begin{itemize}
\item \underline{ Coulomb dissociation.} One simply gets
\be\label{eq:ampl1}
\begin{split}
F^{\mathrm c}_\mathrm{dis}=&
\frac{ip_d}{2\pi}G\left(\textstyle{\frac12}\mathbf{ Q, k}\right)\times\\
&\times\int d^2b_pe^{i\mathbf b_p\mathbf Q}\left[e^{-i\chi_\mathrm{scr}}-e^{i\chi_\mathrm{c}(b_p)}\right],
\end{split}
\ee
where
\be\label{eq:ffactor}
G\left(\textstyle{\frac12}\mathbf{ Q, k}\right)=
\int d^3 r e^{\frac{i}2 \mathbf{Q r}} \psi^\ast_\mathbf{\,k}(\mathbf r)\psi_0(\mathbf r)
\ee
is the transition form factor. To perform integration over $d^2b_p$ it is useful to introduce  a point charge phase shift \cite{GlauberMatthiae}
\be\label{eq:3}
\chi_\mathrm{pt}(b_p)=\frac{2 Z \alpha}{v_p}\ln p_3 b_p= \chi_0+\widetilde\chi_\mathrm{pt}(b_p),
\ee
where
\[\widetilde\chi_\mathrm{pt}(b_p)=\frac{2 Z \alpha}{v_p}\ln b_p,\]
and rewrite identically
$
\left[e^{-i\chi_\mathrm{scr}}-e^{i\chi_\mathrm{c}}\right]=\left[e^{-i\chi_\mathrm{scr}}-e^{i\chi_\mathrm{pt}}\right]+\left[e^{i\chi_\mathrm{pt}}-e^{i\chi_\mathrm{c}}\right]$.
Finally one arrives at
\be
\begin{split}
F^{\mathrm c}_\mathrm{dis}=&G\left(\textstyle{\frac12}\mathbf{ Q, k}\right)[\mathcal F_\mathrm{pt}(Q) + \Delta\mathcal F(Q)]\equiv\\
\equiv& G\left(\textstyle{\frac12}\mathbf{ Q, k}\right)\mathcal F_\mathrm{c}(Q) ,
\end{split}
\ee
where
\be \label{Coulob_pot}
\mathcal F_\mathrm{pt}(Q)=-\frac{2Z\alpha p_d}{vQ^2}e^{i\varphi_\mathrm{c}}
\ee
is the scattering amplitude on Coulomb potential of a point charge ($\varphi_\mathrm{c}=-\frac{2Z\alpha}{v}\left[C+\ln\left(\frac{Q}{2p_3}\right)\right]$, where $C=0.577215$ is Euler constant) and 
\be
\begin{split}
\Delta \mathcal F(Q) &\equiv e^{i\chi_0}\Delta \mathcal{\widetilde F}(Q)=\\
&=ip_d\int_0^\infty db_p b_p J_0(Qb_p)\left[e^{i\chi_\mathrm{pt}(b_p)}-e^{ i\chi_\mathrm{c}(b_p)}\right],
\end{split}
 \label{Correction}
\ee
where $J_0(x)$ is the Bessel function.
\item \underline{Proton scattering affected by Coulomb interaction.} 
%
%
%
\be
F^{\mathrm{\;c}}_p
=\frac{\sigma}{4\pi}G\left(-\textstyle{\frac12}\mathbf{ Q, k}\right)\mathcal C_c(Q), 
\ee
where $\mathcal C_c(Q)=\int_0^\infty dq q e^{-\frac{\beta^2}2 (q^2+Q^2)}I_0(\beta^2 Qq)\mathcal F_\mathrm{c}(q)$.
\item \underline{Neutron scattering affected by Coulomb interaction.} 
\be
F^{\mathrm{\;c}}_n=
\int \frac{d^2 l}{2\pi in} G\left(\textstyle{\frac12}\mathbf{Q-l, k}\right)f_n(l)\mathcal F_\mathrm{c}(|\mathbf Q_\perp -\mathbf l_\perp|).
\ee
%
%
%
%
\item \underline{$n-p$ rescattering affected by Coulomb interaction.}
\be
\begin{split}
F^{\mathrm{\;c}}_{np}=&
-\frac1{(2\pi)^2pn}\int d^2l d^2l' G\left(-\textstyle{\frac12} \mathbf Q + {\mathbf l}',\mathbf k\right) \times\\
&\times \mathcal F_\mathrm{c}(|\mathbf{Q-l-l'}|)f_p(l)f_n(l').
\end{split}
\ee
For gaussian amplitudes $f_p(l)$ and $f_n(l')$ one can integrate over the angles
\be
\begin{split}
F^{\mathrm{\;c}}_{np}=&
-\left(\frac{\sigma}{4\pi}\right)^2\int_0^\infty dq \int_0^\infty dq' q q' G\left(\textstyle{\frac12}q',\mathbf k\right)\times \\
&\times  \mathcal F_\mathrm{c}(q)\widetilde I_0\left(-\textstyle{\frac12}\beta^2 qq'\right)\widetilde I_0\left(\textstyle{\frac12}\beta^2 qQ\right)\times \\
&\times  e^{-\frac14\beta^2[(q-q')^2+(q-Q)^2)},
\end{split}\label{DnpC}
\ee
where $\widetilde I_0(x)=e^{-|x|} I_0(x)$. Deriving (\ref{DnpC}) we used the fact that the form factor $G(\mathbf Q',\mathbf k)$ does not depend on the $\mathbf Q$-direction when the vector $\mathbf k$ is directed over $z$ axis. We have also assumed that the nucleon-nuclei amplitudes are the same for the proton and the neutron.
\end{itemize}
Finally one gets
\be
\begin{split}
&F(\mathbf p_\perp,p_3,\mathbf Q_\perp)= \\
&=F^{\mathrm{str.}}+e^{i\chi_\mathrm{scr}}\left(F^{\mathrm{\;c}}_\mathrm{dis} -F^{\mathrm{\;c}}_{p} -F^{\mathrm{\;c}}_{n} + F^{\mathrm{\;c}}_{np}
\right).
\end{split}
\label{AMPL}
\ee

As the result of infinite range of the Coulomb interaction, the amplitude (\ref{Coulob_pot}) is diverged at $Q \to 0$. Due to the deuteron bound energy the neutron loses a part of its longitudinal momentum $Q_0$ at the limit when the transverse momentum $Q_\perp \to 0$ \cite{AkhiezerSitenko}. We estimate it to be
\be
\label{CS3}
Q_0=\frac{m_n^2+(p_d-p)^2-(E_d-E_p)^2}{2(p_d-p)}.
\ee
\begin{figure}[b]
\centering
\includegraphics[width=0.5\textwidth]{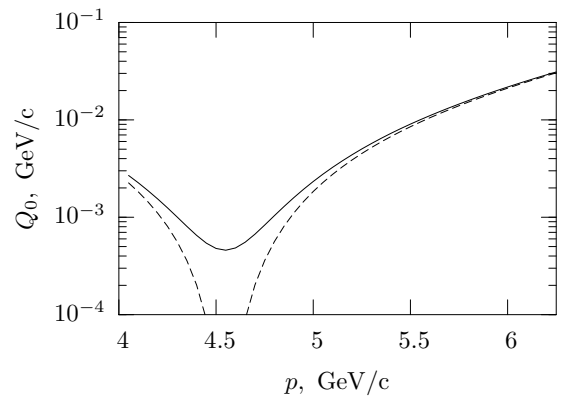}
\caption{Dependence of the $Q_0$ momentum (\ref{CS3}) on the proton longitudinal momentum $p$ for the deuteron momentum $p_d=9.1$~GeV/c. The solid line is for the experimental masses of the proton and the neutron and the dashed line is for $m_p=m_n=\frac12m_d$.}
\label{fig:oscillation}
\end{figure}
To take into account this effect, one has to change $Q^2 \to Q^2 +Q_0^2$ in the denominator of the amplitude (\ref{Coulob_pot}). The momentum $Q_0$ is minimal at $p=\frac12 p_d$ (see Figure~1) what means that at this kinematical region  the Coulomb interaction should be maximal
.
\subsection{Strong interaction part of the amplitude \label{sec:1.2}}
As was mentioned before, the strong interacting amplitude (\ref{eq:1}) corresponds to the situation when the deuteron constituents (the proton and the neutron) suffer elastic scattering only. Its square is proportional to the so-called ``disintegration cross section''. To calculate the inclusive cross section, one has to add two contributions, the  disintegration and absorption cross sections \cite{BTekou,BTreleani}. In the later the neutron suffers inelastic collisions, but the proton keeps elastic scattering only. It is the core of Bertocchi-Treleani model.  

In Ref.~\cite{Kobushkin98} Bertocchi-Treleani model \cite{BTreleani} was modified in the following way:
\begin{itemize}
\item the deuteron wave function was considered in the framework of ``minimal relativization prescription'' with dynamics in the infinite momentum frame \cite{FStrikman,KobViz};
\item it takes into account Pauli principle at the constituent quark level.
\end{itemize}
According to Resonating Group Method (RGM) Pauli principle at the level of constituent quarks modifies the deuteron wave function at short distances by \cite{GKS,GlozmanKuchina}
\be
\psi^d(1,2,...,6)=\widehat A {\phi_N(1,2,3)\phi_N(4,5,6)\chi(\mathbf r)},
\label{RGMWF}
\ee
where $\widehat A$ is the antisymmetrizer for quarks from different three quark ($3q$) bags, $\phi_N$ are nucleon $3q$ clusters, $\chi(\mathbf r)$ is the RGM distribution function and $\mathbf r$ stands for the relative coordinate between two $3q$ bags. Due to the presence of the antisymmetrizer in (\ref{RGMWF}), the deuteron wave function, being decomposed into $3q \times 3q$ clusters, includes, apart from the standard $pn$ component, nontrivial $\mathrm{NN}^\ast$ components. Most of the isobars $\mathrm{N}^\ast$ have  negative parity and thus they generate effective $P$-wave in the deuteron. 

Following Refs.~\cite{GKS,GlozmanKuchina} one can choose $\chi(\mathbf r)$ as a conventional two nucleon deuteron wave function renormalized by the condition, formulated in Ref.~\cite{OkaYazaki}.
\subsection{The cross section \label{sec:1.3}}
The differential cross section is given by
\be
\label{CS1}
\frac{d^3\sigma}{d^3 k}=\frac1{(2\pi)^3}\int d^2n
\left|F(\mathbf p_\perp,p_3,\mathbf Q_\perp)\right|^2,
\ee
where $\mathbf p$ and $\mathbf n$ are the proton and neutron momenta. In the case, when $p_\perp=0$ the transverse momentum $\mathbf Q=\mathbf n_\perp$ and the integral over the angle becomes trivial. Finally one arrives at the following expression for the invariant differential cross section
\be
\label{CS2}
E_p\frac{d^3\sigma}{d^3 p}=\frac{E_p^\ast}{(2\pi)^2}\int_0^\infty dnn
\left|F(\mathbf 0_\perp,p_3,n)\right|^2,
\ee
where $E_p^\ast$ is the proton energy in the deuteron rest frame.

Due to oscillation factor $e^{i\chi_\mathrm{scr}}$ in (\ref{AMPL}), the strong interacting and Coulomb parts of the amplitude do not interfere in (\ref{CS2}).
\section{Numerical calculations \label{sec:numerical}}
We use the nuclear charge density which corresponds to the harmonic oscillator well
\be\label{NUM1}
\rho_\mathrm{c}(r)=\frac{2}{\pi^\frac32 Z a_0^3}\left[1+(Z-2)\frac{r^2}{3a^2_0}\right]e^{-r^2/a^2_0},
\ee
where $\langle r^2 \rangle=a_0^2\left(\frac52-\frac{A}4\right)$. For $^{12}$C the parameter $a_0=1.60$~fm \cite{GlauberMatthiae}. The thickness function
\be\label{NUM2}
T_\mathrm{c}(b)=\frac2{\pi Z a_0^2}\left[1+\frac{Z-2}6 +(Z-2)\frac{b^2}{3a_0^2}\right]e^{-b^2/a_0^2}.
\ee

The Coulomb phase shift $\widetilde\chi_\mathrm{c}(b_p)$ is displayed at Figure~2.
\begin{figure}[b]
\centering
\includegraphics[width=0.5\textwidth]{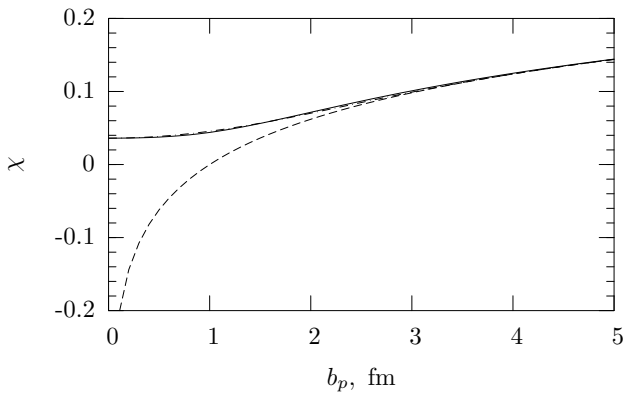}
\caption{The Coulomb phase shift $\widetilde\chi_\mathrm{c}(b_p)$ (the solid line) and the phase shift for a point charge $\widetilde\chi_\mathrm{pt}(b_p)$ (the dashed line) for $^{12}$C. The dashed-point line is for the approximation of the phase shift $\widetilde\chi_{\mathrm{ap.}}(b_p)$.}
\end{figure}

In numerical calculations we use parameterizations for the phase shift $\widetilde\chi_\mathrm{c}(b)$ (\ref{eq:2}) and the real and imaginary parts of the potential correction $\Delta\mathcal{\widetilde F}_\mathrm{c}(Q)$ defined in Eq.~(\ref{Correction}) which are given in Appendix.

All calculations are done with the deuteron wave function for Nijm-I potential \cite{NIJM}. The $S$ and $D$ wave components of the wave function were approximated by sum of gaussians
\be
u(r)=r\sum_{i=1}^N A_ie^{-\alpha_i r^2},\quad w(r)=r^3\sum_{i=1}^N B_ie^{-\beta_i r^2}.
\label{WF}
\ee
For the gaussian wave function (\ref{WF}) one cannot construct wave function for the unbound $pn$ system which fulfill the conditions of orthogonality and complitnes. Similar to \cite{EPS} one can construct only a function which is orthogonal to the deuteron wave function (\ref{WF})
\be
\psi_{\mathbf k}(\mathbf r)=e^{i\mathbf{kr}}-(2\pi)^{3/2}\psi_s(r)\phi_s(k)/N_s,
\ee
where $\psi_s(r)=\frac1{(4\pi)^{1/2}r}u(r)$, $N_s$ is probability for the $S$-wave in the deuteron and $\phi_s(k)$ is the Fourier transform of the deuteron $S$-wave function, 
\[\phi_s(k)=\sqrt{\frac1{4\pi}}\sum_{i=1}^N \frac{A_i}{(2\alpha_i)^{3/2}}e^{-k^2/(4\alpha_i)}.\] Finally one gets the following expression for the form factor
\be
\begin{split}
&G\left(\textstyle{\frac12}\mathbf Q,\mathbf k\right)=(2\pi)^{3/2}\left\{
\phi_s\left(\left|\textstyle{\frac12}\mathbf{Q-k}\right|\right)-\right.\\
&\left.
-\pi^{3/2}\frac{\phi_s(k)}{N_s}\sum_{i,j=1}^N \frac{A_iA_j}{(\alpha_i+\alpha_j)^{3/2}}\exp\left[-\frac{Q^2}{16(\alpha_i+\alpha_j)}\right]
\right\}.
\end{split}
\ee
In numerical calculations we take for the total $p^{12}$C cross section its experimental value $\sigma=$340~mb. The slop parameter was calculated in the framework of Glauber-Sitenko model to be $\beta^2=69.3\ \mathrm{(GeV/c)^2}$.
\begin{figure}[t]
\centering
\includegraphics[width=0.5\textwidth]{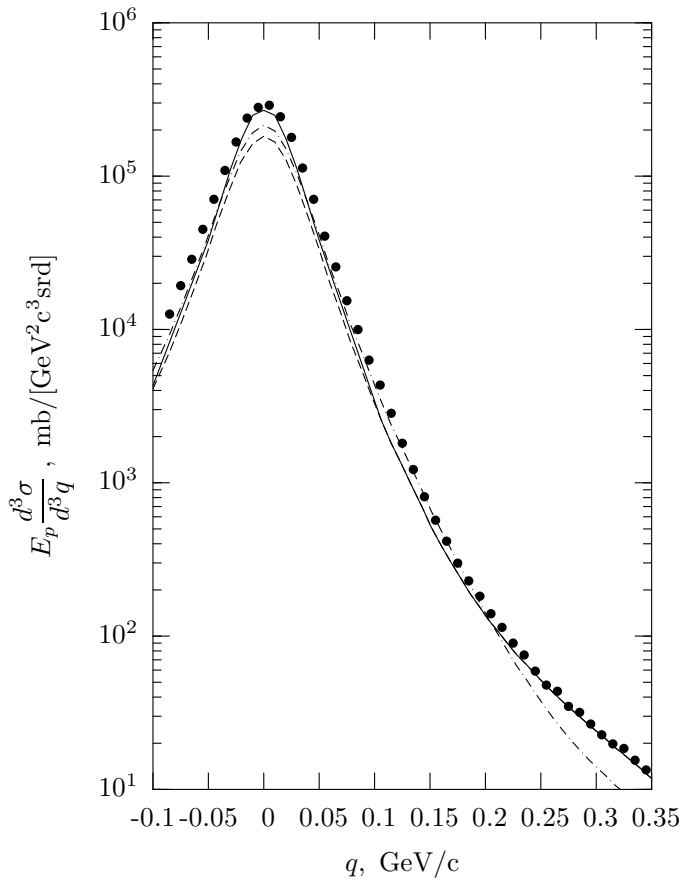}
\caption{The invariant cross section of the $0^\circ$ inclusive $^{12}\mathrm C(d,p)$ break-up at $p_d=9.1$~GeV/c plotted versus the proton momentum in the deuteron rest frame. The experimental points are from \cite{BUPCS1,BUPCS2}.  The dash-dotted curve is for the quasi-impulse approximation (see text), the dashed curve is for multiple scattering~+~Pauli principle at quark level (PQL), the full curve is for  multiple scattering~+~PQL corrected by Coulomb interaction.}
\label{fig:CS}
\end{figure}
\section{Comparison with experiment and summary \label{sec:summ}}
In Figure~3 the results of calculations are compared with experimental cross section data. We also compare our results with quasi-impulse approximation, i.e. $\frac{d^3\sigma}{d^3q}\approx \sigma^\mathrm{in}_{n\mathrm C} |\psi_d(q)|^2$, where $q$ is the proton momentum in the deuteron rest frame and $\sigma^\mathrm{in}_{n\mathrm C}$ is the inelastic neutron-carbon cross section\footnote{For the meaning of the quasi-impulse approximation see, e.g. Ref.~\cite{BUPCS1}.}. One sees that the Coulomb interaction strongly inreases the cross section at the peak near $q\sim 0$ (which corresponds to $p\sim \frac12 p_d$). Out of the peak such effect sharply decreases and becomes negligibly small at $q>$100 MeV/c. So it cannot affect the region, $q>$200~MeV/c, where quark effects are assumed to be significant.

\vspace{0.5cm}
The authors would like to thank E.A.~Strokovsky for reading manuscript and useful comments. This work was supported by Grant of State Foundation of Fundamental Research of Ukraine F/16-457-2007.
\appendix
\section{}
In numerical calculations we use the following parameterizations for the phase shift $\widetilde\chi_\mathrm{c}(b)$
\be\label{NUM3}
\widetilde\chi_{\mathrm{ap.}}(b)=\frac{Z\alpha}{v_p}\ln\left(b^2+\frac{A_1}{1+A_2b^2}\right)
\ee
(where $A_1=$2.2416~fm$^2$, $A_2=$0.34~fm$^{-2}$ and $b$ is defined in fm) and  the real and imaginary parts of the potential correction $\Delta\mathcal{\widetilde F}_\mathrm{c}(Q)$ 
\be\label{NUMREIM}
\begin{split}
&\Re\mathrm e\Delta\mathcal{\widetilde F}_\mathrm{c}(Q)=-\frac{2Z\alpha p_d}{v}\left[\frac{\cos \varphi_r}{A_r+Q^2} + C_r Q e^{-\frac{Q^2}{Q_r^2}}\right],\\
& \Im\mathrm m \Delta \mathcal{\widetilde F}_\mathrm{c}(Q)=-\frac{2Z\alpha p_d}{v}\left[\frac{A_i}{B_i+Q^2}+C_ie^{-\frac{Q^2}{Q^2_i}}\right],
\end{split}
\ee
where $\varphi_r=-\frac{2Z\alpha p_d}{v}\left[C+\ln\left(\frac{B_r+Q}{2p_3}\right)\right]$ and $A_r=0.035\ \mathrm{GeV}^2$, $B_r=0.086\ \mathrm{GeV}$, $C_r=-12.410\ \mathrm{GeV}^{-3}$, $Q^2_r=0.125\ \mathrm{GeV}^2$, $A_i=0.264$, $B_i=0.799\ \mathrm{GeV}^2$, $Q^2_i=0.0242\ \mathrm{GeV}^2$, $C_i=-1.112\ \mathrm{GeV}^{-2}$. 
%


\begin{thebibliography}{99}
\bibitem{BUPCS1}
{\it Ableev V.G. et al.} //Nucl. Phys. -- 1983. -- {\bf A 393}. -- P. 491-501; ibid -- 1984. -- {\bf A 411-501}. -- P. 591 (E); Pis'ma v ZhETF -- 1983. -- {\bf 37}. -- P. 196-198; JINR Rap. Comm. -- 1992. -- {\bf 1[54]}. -- P. 10.
\bibitem{BUPCS2} 
{\it Zaporozhets S.A. et al.} //Proceedings of VIII International Seminar in High Energy Problems (Dubna, June 1986). -- 1986. -- D1,2-86-668. -- P. 341-349.
\bibitem{BUPT20Perdrisat}
{\it Perdrisat C.F. et al.} //Phys. Rev. Lett. -- 1987. -- {\bf 59}. -- P. 2840-2843.
\bibitem{BUPT20Ableev}
{\it Ableev V.G. et al.} // Pis'ma v ZhETF -- 1988. -- {\bf 47}. -- P. 558-561.
\bibitem{BUPT20Punjabi}
{\it Punjabi V. et al.} // Phys. Rev. -- 1989. -- {\bf C 39}. -- P. 608-618.
\bibitem{BUPT20Ableev90}
{\it Ableev V.G. et al.} //JINR Rap. Comm. -- 1990. -- {\bf 4[43]-90}. -- P. 5.
\bibitem{BUPT20Nomifilov}
{\it Nomofilov A.A. et al.} //Phys. Lett. -- 1994. -- {\bf B 325}. -- P. 327-332.
\bibitem{BUPT20Aono}
{\it Aono T. et al.} //Phys. Rev. Lett. -- 1995. -- {\bf 74}. -- P. 4997-5000.
\bibitem{BUPT20Azgirey}
{\it Azgirey L.S. et al.} //Phys. Lett. -- 1996. -- {\bf B 387}. -- P. 37.
\bibitem{BUPKAPPACheunh}
{\it Cheung N.E. et al.} //Phys. Lett. -- 1992. -- {\bf B 284}. -- P. 210-214.
\bibitem{BUPKAPPADzikowski} 
{\it Dzikowski T. et al.} //Proceedings of International Seminar in Workshop `Dubna 97', JINR. -- 1992. -- E2-92-25. -- P. 181.
\bibitem{BUPKAPPAKuehn}
{\it Kuehn B. et al.} //Phys. Lett. -- 1994. -- {\bf B 334}. -- P. 298;
{\it Azgirey L.S. et al.} //JINR Rap. Comm. -- {\bf 3[77]-96}. -- P. 23.
\bibitem{EBSCS}
{\it Berthet P. et al.} //J. Phys. G:Nucl. Phys. -- 1982 -- {\bf 8}. -- P. L111. 
\bibitem{EBST20}{\it Azgirey L.S. et al.} //Phys. Atom. Nucl. -- 1998. -- {\bf 61}. -- P. 432-447 [Yad. Fiz.  -- 1998. -- {\bf 61}. -- P. 494-510]; Phys. Lett. -- 1997. -- {\bf  B 391}. -- P. 22 -28. 
\bibitem{EBSKAPPA}
{\it Punjabi~V. et al.} //Phys. Lett. -- 1995. -- {\bf B350}. -- P. 178-183.
\bibitem{Kobushkin98}
{\it Kobushkin~A.P.} //Phys. Lett. -- 1998 -- {\bf B 421}. -- P. 53-58; Phys. Atom. Nucl. -- 1999. -- {\bf 62}. -- P. 1400-1146 [Yad. Fiz. -- 1999. -- {\bf 62}. -- P. 1213-1219]; A. P. Kobushkin, in Proceedings of the RCNP-TMU Symposium ``Spins in Nuclear and Hadronic Reactions'', ed. by H. Yabu, T. Suzuki, and H. Toki (Word Sci., Singapore, 2000) p. 223.
\bibitem{GKS}
{\it Glozman~L.Ya., Kobushkin~A.P., and Syamtomov~A.I.} //Phys. Atom. Nucl. -- 1996. -- {\bf 59}. -- P. 795-803 [Yad. Fiz. -- 1996. --{\bf 59}. -- P. 833-841].
\bibitem{GlozmanKuchina}
{\it Glozman~L.Ya. and Kuchina~E.I.} //Phys. Rev. -- 1994. -- {\bf C 49}. -- P. 1149-1165.
\bibitem{AkhiezerSitenko}
{\it Akhiezer~A.I. and Sitenko~A.G.} //Phys. Rev. -- 1957. -- {\bf 106}. -- P. 1236.
\bibitem{EPS}
{\it Evlanov~M.V., Polozov~A.D., and Struzhko~B.G.} //Ukr. P. J. -- 1980. -- {\bf 25} -- P. 813.
\bibitem {GlauberMatthiae}
{\it Glauber~R.J. and Matthiae~G.} //Nucl. Phys. -- 1970. -- {\bf B21}. -- P. 135-157.
\bibitem{BTekou}
{\it Bertocchi~L. and T\'ekou~A.} //Nuovo Cim. -- 1974. -- {\bf 21 A}. -- P. 223.
\bibitem{BTreleani}
{\it Bertocchi~L. and Treleani~D.} //Nuovo Cim. -- 1976. -- {\bf 36 A}. -- P. 1-22.
\bibitem{FStrikman}
{\it Frankfurt~L.L. and Strikman~M.I.} // Yad. Fiz. -- 1979. -- {\bf 29}. -- P. 490; Phys. Rep. -- 1976. -- {\bf 76}. -- P. 215.
\bibitem{KobViz}
{\it Kobushkin~A.P. and Vizireva~L.} //J. Phys. G:Nucl. Phys. -- 1982. -- {\bf 8}. -- P. 893.
\bibitem{OkaYazaki}
{\it Oka~M. and Yazaki~K.} //Prog. Theor. Phys. -- 1981. -- {\bf 66}. -- P. 556.
\bibitem{NIJM}
{\it Stoks~W.G., Klomp~R.A.M., Terheggen~C.F.P., and de~Swart~J.J.} //Phys. Rev. -- 1994. -- {\bf C 49}. -- P. 2950.
\end{thebibliography}
\end{document}